\pdfoutput=1
\documentclass[a4paper]{article}
\usepackage{amsfonts}
\usepackage{amssymb}
\usepackage{amsmath}
\usepackage{graphicx}
\usepackage{authblk}
\usepackage{bbold}

\newcommand{\ket}[1]{| #1 \rangle}
\newcommand{\bra}[1]{\langle #1 |}
\newcommand{\ketbra}[2]{| #1 \rangle \langle #2 |}
\newcommand{\braket}[2]{\bra{#2}{#1}\rangle}
\newcommand{\tr}{\mathrm{tr}}
\newcommand{\1}{\mathbb{1}}

\newtheorem{prop}{Proposition}

\newtheorem{definition}{Definition}
\newtheorem{exam}{Example}

\newcommand{\stackidx}[4]{
  \substack{
  #1  #2 \\
  #3  #4}
}

\title{A model for quantum queue}
\author[$*$]{Piotr Gawron}
\author[$*$,$\dagger$]{Dariusz Kurzyk}
\author[$*$,$\ddagger$]{Zbigniew Pucha\l{}a}
\affil[$*$]{Institute of Theoretical and Applied Informatics, Polish Academy
of Sciences, Ba{\l}tycka 5, 44-100 Gliwice, Poland}
\affil[$\dagger$]{Institute of Mathematics, Silesian University of Technology, 
Kaszubska 23, Gliwice 44-100, Poland}
\affil[$\ddagger$]{Instytut Fizyki im. Smo\l{}uchowskiego, Uniwersytet 
Jagiello\'nski Reymonta 4, 30-059 Krak\'ow, Poland}

\date{October 31, 2012}

\begin{document}
\maketitle
\begin{abstract}
We consider an extension of Discrete Time Markov Chain queueing model to the
quantum domain by use of Discrete Time Quantum Markov Chain. We introduce
methods for numerical analysis of such models. Using this tools we show that
quantum model behaves fundamentally differently from the classical one.
\end{abstract}
\section{Introduction}

Today the most important application field for queuing theory is
telecommunication. Queuing theory applied to description of packet switching in
network routers allows to foresee packets behaviour and determine the
distribution of waiting time for transmission and probability of loosing packets
caused by buffers overflows. This allows to anticipate the quality of
telecommunication services. As variability of flows is an important factor of
delays and losses, it is difficult to obtain exact analytical solutions for wide
spectrum of models. Markov chains is typical method applied to analysis of 
transitient states of queuing systems.

In this paper we introduce a new approach to queuing networks based on quantum
information theory. We consider simple models of queuing systems described by 
Discrete Time Quantum Markov Chains (DTQMC) which are reducible to Discrete 
Time Markov Chains (DTMC). Our model is an adaptation of quantum random walks 
(QRW) where both walker and the system are described by quantum channels. It 
will be shown that quantum queueing systems behave differently than classical 
ones.

Discrete Time Quantum Markov Chains are a~tool which is used in the theory of 
quantum information for example to model noisy behaviour in quantum iterative 
algorithms \cite{gawron2012noise}.

We want this paper to be self-consistent therefore in
Section~\ref{sec:formalism} we recall essential basic notions from quantum
information theory. The rest of the paper is organised as follows, in
Section~\ref{sec:qunatum-queue} we introduce the model of a quantum queue, in
Section~\ref{sec:tools} we gather the mathematical tool essential for analysis
of DTQMCs, in Section~\ref{sec:stability} we discuss the problem of steady state
distribution for quantum queue, in Section~\ref{sec:classical} we show that our
model reduces to classical one under some constraints, in
Section~\ref{sec:mean-distribution} we analyse quantum queues initialized with
random initial states, and in Section~\ref{sec:conclusions} we draw final
conclusions.
\section{Formalism of quantum information}
\label{sec:formalism}

\subsection{Dirac notation}
Throughout this paper we use Dirac notation. Symbol $\ket{\psi}$ denotes
a~complex column vector, $\bra{\psi}$ denotes the row vector dual to
$\ket{\psi}$. The scalar product of vectors $\ket{\psi}$, $\ket{\phi}$ is
denoted by $\braket{\phi}{\psi}$. The outer product  of these vectors is denoted
as $\ketbra{\phi}{\psi}.$ Vectors are labelled in the natural way: 
$\ket{0}:=\left(1,0\right)^T$, 
$\ket{1}:=\left(0,1\right)^T$. 
Notation like
$\ket{\phi\psi}$ denotes the tensor product of vectors and is equivalent to
$\ket{\phi}\otimes\ket{\psi}$.

\subsection{Density operators}

The most general state of a~quantum system is described by a~density operator.
In quantum mechanics a~density operator $\rho$ is defined as hermitian
($\rho=\rho^\dagger$), positive semi-definite ($\rho\geq 0$), trace one
($\tr{(\rho)}=1$) operator. When a basis is fixed the density operator can be
written in the form of a~matrix. Diagonal density matrices can be identified
with probability distributions, therefore this formalism is a~natural extension
of probability theory. 

Density operators are usually called quantum states. The set of quantum states
is convex \cite{BengtssonZyczkowski2008} and its boundary consists of pure
states which in matrix terms are rank one projectors. Convex combinations of
pure states lie inside the set and are called mixed states.

\subsubsection{Entanglement}

Entanglement is one of the most important phenomena in quantum information
theory.  We say that state $\rho$ is separable iff it can be written in the
following form
\begin{equation}
\rho=\sum_{i=1}^{M} q_i\, \rho_i^A \otimes \rho_i^B,
\end{equation}
where $q_i>0$ and $\sum_{i=1}^{M} q_i=1$.
A state that is not separable is called entangled. 

\subsubsection{Subsystems}

Given two states $\rho^A$, $\rho^B$ of two systems $A$ and $B$, the product
state $\rho^{AB}$ of the composed system is obtained by taking the Kronecker
product of the states i.e. $\rho^{AB}=\rho^A\otimes\rho^B.$

Let $[\rho^{AB}]_{kl}$ be a matrix representing a~quantum system composed of two
subsystems of dimensions $M$ and $N$. We want to index the matrix elements of
$\rho$ using two double indices $[\rho^{AB}]_{\stackidx{m}{\mu}{n}{\nu}},$ so
that Latin indices correspond to the system $A$ and Greek indices correspond to
the system $B$. The relation between indices is as follows $k=(m-1) M + \mu$,
$l=(n-1) N + \nu$.  The partial trace with respect to system $B$ reads
$\rho^A:=\tr_B(\rho^{AB})=\sum_\mu \rho_{\stackidx{m}{\mu}{n}{\mu}}$,  and the
partial trace with respect to system $A$ reads $\rho^B:=\tr_A(\rho^{AB})=\sum_m
\rho_{\stackidx{m}{\mu}{m}{\nu}}$.

Given the state of the composed system $\rho^{AB}$ the state of subsystems can
by found by the means of taking partial trace of $\rho^{AB}$ with respect to one
of the subsystems. It should noted that tracing-out is not a reversible
operation, so in a~general case
\begin{equation}
    \rho^{AB}\neq \tr_A(\rho^{AB})\otimes\tr_B(\rho^{AB}).
\end{equation}

\subsection{Completely positive trace-preserving maps (CPTP)}

We~say that an operation is physical if it transforms density operators into
density operators. Additionally we assume that physical operations are linear.
Therefore an operation $\Phi(\cdot)$ to be physical has to fulfil the following
set of conditions:
\begin{enumerate}
	\item For any positive operator $\rho$ its image under operation $\Phi$ has
	to have its trace and positivity preserved i.e. if 
	$\rho\geq 0$ then $\Phi(\rho)\geq 0$ and $\tr{(\Phi(\rho))} = \tr \rho.$
	\item Operator $\Phi$ has to be linear:
	\begin{equation}
		\Phi\left(\sum_i p_i\rho_i \right)=\sum_i p_i \Phi\left(\rho_i \right).
	\end{equation}
	\item The extension of the operator $\Phi$ to any larger dimension that acts 
	trivially on the extended system has to preserve positivity. This feature is
	called complete positivity. Extended channel $(\Phi\otimes\1_{K})$ acting on
	operators on $\mathcal{H}_N \otimes \mathcal{H}_K$ is
	defined for the product states as 
	\begin{equation}
		(\Phi\otimes\1_{K})\left(\rho\otimes\xi\right) =\Phi\left(\rho\right)\otimes\xi
	\end{equation}
	and extended for all states by linearity.
	Thus, the complete positivity means, that for a positive operator $\sigma$
	acting on $\mathcal{H}_N \otimes \mathcal{H}_K$, we have 
	\begin{equation}
		(\Phi\otimes\1_{K})(\sigma) \geq 0.
	\end{equation}
\end{enumerate}
CPTP maps are often called quantum channels~\cite{BengtssonZyczkowski2008,bruzda2009random,puchala2011experimentally}.

Any operator $\Phi$ that is completely positive and trace preserving can be
expressed in so called Kraus form \cite{BengtssonZyczkowski2008}, which consists
of the finite set $\{E_k\}$ of Kraus operators -- matrices that fulfil the
completeness relation: $\sum_k {E_k}^\dagger E_k=\1$. The image of state $\rho$
under the map $\Phi$ is given by 
\begin{equation}
    \Phi(\rho)=\sum_k E_k \rho {E_k}^\dagger.
\end{equation}

\subsection{Quantum measurement}

The most general approach to the quantum measurement is given by \emph{Positive
Operator Valued Measures} (POVM) \cite{BengtssonZyczkowski2008}. Any partition
of identity operator into a set of $N$ positive operators $F=\{F_i\}_{i=1}^N$
such that
\begin{equation}
\sum_i^N F_i=\1
\end{equation}
and set of real measurement outcomes $O=\{o_i\}_{i=1}^N$ together with a 
mapping $\mu:O\to F$ is called POVM measurement. The probability
$p_i$ associated with any possible measurement outcome $o_i$ is given by $p_i=\tr(F_i\rho)$, where $\mu(o_i)=F_i$.

\section{Quantum queue}
\label{sec:qunatum-queue}
The model of a~quantum queue presented in this work is loosely based on models
of discrete time quantum random walks (QRW). Discrete time quantum random  walks
are usually realized by two coupled quantum systems: quantum coin and walker.
The evolution of this kinds of systems is governed by application of a unitary
operator on the coin subsystem with application of conditional operation on the
both systems such that only the state of the coin influences the state of the
walker. For a~elementary introduction to the subject of QRW please refer to
\cite{Kempe2003}.

In our terms we split the coin into two quantum coins which represent the flow
of jobs into and out of the queueing system. The walker represents the length of
the queue. We assume that capacity of the queue is finite and number of jobs in
the queue cannot drop below zero therefore our model might be compare to QRW
with barriers \cite{BachEtAll2004}.

A quantum queue is modelled in the following way. We chose a quantum system
consisting of three subsystems which we will identify with input $I$, output $O$
and queue $Q$. We will apply repetitively CPTP maps on our quantum systems 
effectively forming discrete time quantum Markov chain and measure the state of 
subsystem $Q$.

\subsection{Transition}
\begin{figure}[h]
    \begin{center}
        \includegraphics[width=62mm]{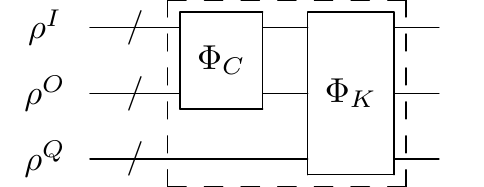}
    \end{center}
\caption{Schematic representation quantum queue transition.}\label{fig:queue}
\end{figure}

Flow of jobs into the queue, flow of the processed jobs filling and emptying of
the queue are described by quantum channels. We define two quantum channels:
$\Phi_C$ that represents the input and output coins and  $\Phi_K$ that
represents the conditional change of the queueing system state. Schematic
depiction in form of the quantum circuit of the system is presented in
Fig.~\ref{fig:queue}.

\subsection{Queue action}

Let $D_{I}, D_{O}$ are the dimensions of registers for jobs that flow into
system in a single time step, the jobs processed in a single time step,
respectively and $D_{Q}$ be dimension of the register representing queue length.

We define set of Kraus operators that model the protocol governing the action of
the quantum queue
\begin{eqnarray}
\label{eqn-krauss-S}
K_s&=&\sum_{n=0}^{D_{I}-1} \sum_{m=0}^{D_{O}-1} \sum_{j=D_{O}-1}^{D_{Q}-D_{I}} 
\ketbra{n}{n}\otimes\ketbra{m}{m}\otimes\ketbra{j+n-m}{j},\\
\label{eqn-krauss-L}
K_{l_i}&=&\sum_{n=0}^{D_{I}-1} \sum_{m=0}^{D_{O}-1} 
\ketbra{n}{n}\otimes\ketbra{m}{m}\otimes\ketbra{\max(l_i+n-m,0)}{l_i},\\
\label{eqn-krauss-U}
K_{u_i}&=&\sum_{n=0}^{D_{I}-1} \sum_{m=0}^{D_{O}-1} 
\ketbra{n}{n}\otimes\ketbra{m}{m}\otimes\ketbra{\min(u_i+n-m,D_{Q}-1)}{u_i},
\end{eqnarray}
where $l_i\in L=\{0,1,\ldots,D_{O}-2\}$ and $u_i\in
U=\{D_{Q}-D_{I}+1,D_{Q}-D_{I}+2,\ldots,D_{Q}-1\}$. Set of Kraus operators
$K_{\{s\}\cup L \cup U}$ forms quantum channel $\Phi_K$. 
In order to better understand the
meaning of the above Kraus operators we can divide them into three regimes:
\begin{itemize}
  \item $K_s$ defines evolution of the system in case when the flow of jobs
  can-not cause that state of the queue reaches any of the boundaries. The
  number of jobs waiting in the queue changes by the number of the difference
  between number of jobs flown into and flown out of the queueing system.
  \item Operators $K_{l_i}$ define the behaviour of the queueing system in case
  when flow of jobs may cause reaching of the lower barrier. We can say that
  they ``protect'' the model from reaching the state of negative number of jobs
  waiting in the queue.
  \item Operators $K_{u_i}$ define similar behaviour but for the upper barrier.
\end{itemize}

\subsubsection{Flow of jobs --- coins}

In analogy to quantum random walks we introduce coin operator that defines the
flow of jobs into and from the system.

Coin can be chosen as any quantum channel acting solely on the composite system
$\rho^{I,O}$. One may consider using unitary operators as coins, which is popular
approach in the field of quantum random walks. It should be noted that coins may
form a separable quantum channel on systems $I$ and $O$ or they can act on
compositions of subsystems $I$ and $O$. To be more precise we define a coin as
any CPTP map in the form
\begin{equation}
  \Phi_C(\rho)=\sum_i^N (C_i^{I,O}\otimes \1^Q) \rho (C_i^{I,O}\otimes \1^Q)^\dagger,
\end{equation}
where $\{C_i^{I,O}\}_{i=1}^N$ is any set of Kraus operators.

\subsubsection{Updating of the queue state}

In order to evaluate a single step of queuing system one have to compose 
maps $\Phi_C$, $\Phi_K$ into $\Phi=\Phi_K\circ\Phi_C$.

The evaluation of the system is based on the application of the map $\Phi$ to a
chosen initial quantum  state $\rho_0$, so we form DTQMC.  After $t$ time steps 
the state of the queue becomes
\begin{equation}\label{eq001}
\rho_t=\underbrace{\Phi\circ\Phi\circ\ldots\circ\Phi}_t(\rho_0).
\end{equation}

Probability that the queue is of length $i$ at time $t$  is given by 
\begin{equation}\label{eq:queue-length}
p_t(i)=\bra{i}\rho_t^Q\ket{i},
\end{equation}
where $\rho_t^Q=\tr_{IO}(\rho_t)$ is reduced density matrix.

\section{Tools for analysing quantum queues}
\label{sec:tools}

In this section we gather facts from quantum information theory and theory
of discrete linear dynamical systems which are useful to analysis of quantum
queues.

\subsection{Superoperator}

Kraus representation of quantum channels is sometimes inconvenient to work with.
Quantum channel is a linear map therefore it can be represented in form of the
matrix, which often is called the superoperator. For given set of Kraus
operators $\{E_i\}$ corresponding superoperator $\hat{\Phi}$ is given by the following
relation
\begin{equation}
  \hat{\Phi}=\sum_i^N E_i\otimes E_i^\star,
\end{equation}
where $(\cdot)^\star$ denotes element-wise complex conjugation. It should be 
noted that spectral properties of $\hat{\Phi}$ are important for the analysis of DTQMC.

\subsection{Initial conditions -- random states drawn from HS distribution}\label{randomState}
For initial states in our analysis of quantum queues we take random density
matrices. The theory of random density matrices is a current subject of a much
study~\cite{BengtssonZyczkowski2008,bruzda2009random,puchala2011probability}. In
the  case of random pure states there exists a natural probability distribution,
induced by the Haar measure on the unitary group $U(N)$, called 
\emph{Fubini-Study} measure. However, in some cases, one needs to consider
ensembles of mixed quantum states and then the situation is much more complex.
Firstly because, the probability distribution on the set of density matrices can
be induced using various distance measures. Most commonly used measures are
derived from \emph{Hilbert-Schmidt} or \emph{Bures}
distance~\cite{BengtssonZyczkowski2008}. In this paper we restrict our attention
to the former measure, i.e. \emph{Hilbert-Schmidt measure}, which belongs to the
wider class of probability measures induced by partial trace. 

To obtain a sample form HS distribution we use the following procedure. First we
take a square complex matrix $A$ of size $N$, with real and imaginary part of
each element being \emph{i.i.d.}\! random variables with a standard normal
distribution. Then, we calculate $\rho = \frac{A A^{\dagger}}{\tr A
A^{\dagger}}$. The above, by construction, is~a~density matrix, and moreover is
distributed with HS measure~\cite{BengtssonZyczkowski2008}.

\section{Stability of probability distribution of the queue length}
\label{sec:stability}

The evolution of queuing system for each initial state $\rho_0$ is performed by
a mapping $\Phi$, which is a composition of maps $\Phi_C$, $\Phi_K$. Thus, queue
state of $t$-th step is of the form (\ref{eq001}). Composition the partial trace
over subsystem $Q$ and von Neumann measurement performed over subsystem $Q$
gives probability distribution of the queue length. Denote by $C$ 
a linear operation that traces out subsystems $I$ and $O$ and returns 
probability distribution given by performing measurement over subsystem $Q$ {\sl i.e.}
\begin{equation}
C(\rho)=\sum_i\bra{i}\tr_{IO}(\rho)\ket{i}.
\end{equation}
After $t$ time steps the probability distribution of the queue length is given
by
\begin{equation}\label{eq007}
p_t=C\Phi^{t}(\rho_0).
\end{equation}

In general, superoperator of quantum channel is not hermitian matrix, so its
eigenvalues are in general complex numbers. Since quantum channel $\Phi$ 
transforms convex compact set of density matrices into itself, thus there exists
a fixed point called invariant state $\sigma$, such that $\Phi\sigma=\sigma$. 
This follows from Brouwer fixed point theorem. From above we have that
superoperator $\Phi$  has at least one eigenvalue equal to one, and it can be
shown from the positivity property of the quantum channels, that all eigenvalues
has magnitude less or equal than one \cite{BengtssonZyczkowski2008}.

Superoperator $\Phi$ represents a quantum channel thus the spectrum
$\{z_i\}_{i=1}^N$ of $\Phi$ belongs to the disk $|z_i|\leq 1$. 
Let us order eigenvalues of $\Phi$ according to their  absolute values
$|z_1|\leq |z_2|\leq\cdots \leq |z_N|$. Distribution of queue length depends on
the  moduli of eigenvalues. We can distinguish following cases.

In the first case, if $z_1$ equals to one and $|z_i| < 1$ for $i$ from
$\{2,3,...,N\} $, then the limit $\lim_{t\to\infty}\Phi^t(\rho_0)$ exists  
and is equal to invariant state $\sigma$ for any initial state $\rho_0$
\cite{bruzda2009random}. 

In the second case, if any $|z_j| = 1$ for $j$ from $\{2,3,...,N\}$ then the
limit $\lim_{t\to\infty}\Phi^t(\rho_0)$ might not exist. We are interested
whether the probability distribution of queue length is stable for the operator
$C\Phi^t$, so if there is a limit $\lim_{t\to\infty}p_t$. In order to analyse 
the existence of this limit we will use the notions of linear systems theory.

\begin{definition}[\cite{bernstein1994lyapunov}]
$C\Phi^t$ is semistable if $\lim_{t\to\infty} p_t$ exists for all initial 
conditions $\rho_0$.
\end{definition}

\begin{prop}\label{eq002}
The probability distribution of the queue length is semistable when limit
$\lim_{t\to\infty}(p_{t+1}-p_t)$ exists and coincides to zero vector.
\end{prop}

\noindent Since $p_t$ is a vector from the Hilbert space, hence with the
completeness of the space follows that if there exists the limit
$\lim_{t\to\infty}p_t$ then the limit $\lim_{t\to\infty}(p_{t+1}-p_t)$ coincides
to zero vector.

\begin{prop}\label{eq003}
The probability distribution of the queue length is semistable when 
\begin{equation}\label{eq006}
\lim_{t\to\infty} ||C\Phi^{t+1} - C\Phi^{t}||=0.
\end{equation}
\end{prop}

\noindent Limit from Proposition \ref{eq002} can be written as
$\lim_{t\to\infty} (C\Phi^{t+1} - C\Phi^{t})\rho_0$ and coincides to zero vector
for any initial state $\rho_0$ when $C\Phi^t$ is semistable. Thus if the system
$C\Phi^t$ is semistable then limit (\ref{eq006}) coincides to zero.

The convergence of (\ref{eq007}) can be checked numerical for any $C$ and
$\Phi$. Thus, it is possible to verify if a given model has a stationary
probability distribution of the queue length.

\begin{exam}
Consider an example based on Hadamard coin. Lets fix register dimension of queue length $D_Q$ to $10$, 
thus maximal length of queue can be $9$ and lets set dimensions of registers used to evaluating 
flow of jobs into the system in single time step $D_I=2$ and jobs processed in single 
time step $D_O=2$. We assume that in the initial state the queue is half-filled so it has $5$ jobs.

We set the initial state of the system to
\begin{equation}\label{initialState}
\rho^{IOQ}_0=\frac{1}{2}(\ket{0}-i\ket{1})(\bra{0}-i\bra{1})\otimes
  \frac{1}{2}(\ket{0}-i\ket{1})(\bra{0}-i\bra{1})\otimes
  \ketbra{5}{5}.
\end{equation}

\noindent The coins are unitary and are chosen to be Hadamard gates 
\begin{equation}
H=\frac{1}{\sqrt{2}}
\begin{pmatrix}
1 & 1\\
1 & -1
\end{pmatrix}.
\end{equation}

\noindent After 500 steps, the probability distribution of queue length becomes
stabilized and order of magnitude of $||C\Phi^{500} - C\Phi^{499}||$ is
$10^{-2}$, as shown in the Fig.~\ref{fig:ex0}.

\begin{figure}[h!]
  \begin{center}
    \includegraphics[scale=1]{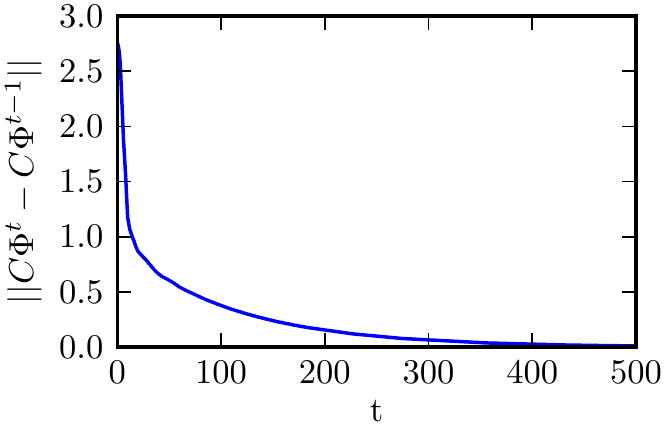}
  \end{center}
  \caption{ Stabilization of the probability distribution of the queue length
  After 500 step, the probability distribution of the queue is stable.}
  \label{fig:ex0}
\end{figure}

The state of the queue reaches lower and upper barriers. In each subsequent time
step the likelihood of barriers increases and the remaining probabilities
converge to zero. The probability distribution of the queue length depends on the
initial state. The choice of half-filled state the queue determines the place of
the middle peak of the likelihood, as shown in the Fig.~\ref{fig:ex1}. 

\begin{figure}[h!]
  \begin{center}
    \includegraphics[width=124mm]{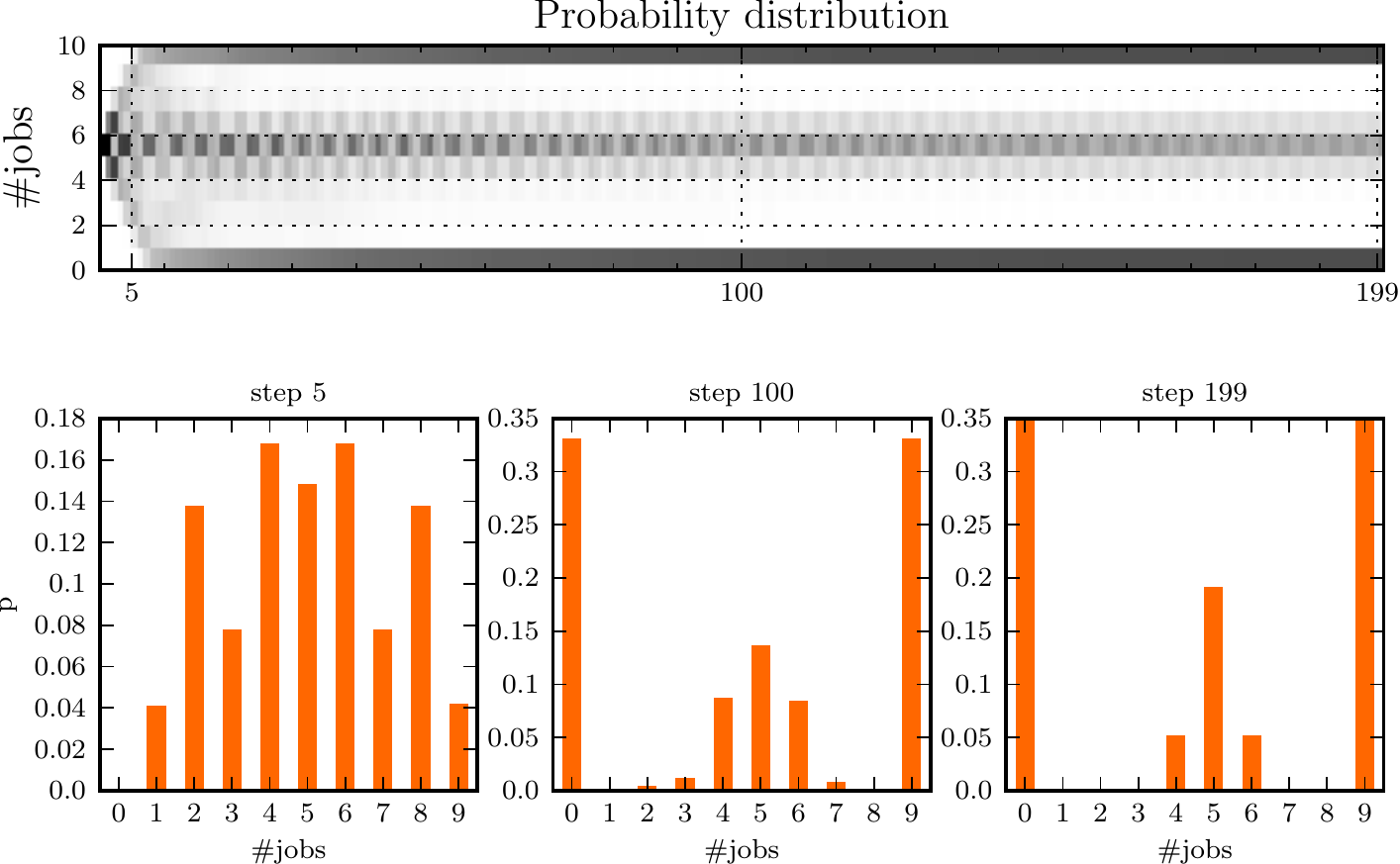}
  \end{center}
  \caption{The probability distribution of the queue length with Hadamard coins. Up to $5$ step, evolution of the quantum queue behaves unitarily. Next, the probability of the queue lengths are distributed at the barriers. The middle peak is the result of choice of the initial state (\ref{initialState}).}
  \label{fig:ex1}
\end{figure}

\end{exam}

\section{Classical system as special case of quantum systems.}
\label{sec:classical}

In a particular case, formalism of discrete time quantum queues can be used to 
describe behaviour of a simple classical queueing system. Classical coin may
be seen as a probability distribution over number of jobs  flowing into the
system and jobs processed. Quantum analogue of classical coin is a quantum
channel that transforms any quantum state into diagonal state. The quantum
description is of course in this case extremely inefficient due to large number
of excessive parameters.

In order to reduce a quantum queuing system to classical one each step of
the evolution a new coin uncorrelated with the state of the queue has to be 
introduced. It can be done by setting $\Phi_C$ to a~composition of two channels:
first one that traces out the coin systems $I$ and $O$, second one that
introduces a coin transforming any quantum state into diagonal quantum state.

Transformation of quantum coin into a classic is performed by a quantum channel
$\Phi_D$ in form
\begin{equation}\label{eq004}
\Phi_D(\rho) = \sum_{x=0}^{D_I\times D_O-1} \ketbra{x}{x}\rho\ketbra{x}{x}.
\end{equation}
Thus classical queue is implemented using a coin of the form
$\Phi_D\cdot\Phi_C$. Consider superoperator corresponding with the channel
$\Phi_D$ given by
\begin{equation}
\hat{\Phi}_D = \sum_{x=0}^{{D_I}^2\times {D_O}^2-1} \ketbra{x}{x}\otimes\ketbra{x}{x}.
\end{equation}
Any superoperator $\hat{\Phi}$ corresponding with quantum channel from
$M$-dimensional space to $L$-dimensional can be presented as 
\begin{equation}
\hat{\Phi} = \sum_{i,j}^M \sum_{k,l}^L c_{i,j,k,l} \ketbra{k}{i}\otimes\ketbra{l}{j},
\end{equation}
where $c_{i,j,k,l}\in\mathbb{C}$ are constants. In our case $M=L=N$.

Composition of $\hat{\Phi}_D$ and any superoperator $\hat{\Phi}$ is in the form
\begin{equation}
\hat{\Phi}_D \cdot\hat{\Phi} = \sum_{x,i,j,k,l}^N c_{i,j,k,l}(\ketbra{x}{x}\otimes\ketbra{x}{x})(\ketbra{k}{i}\otimes\ketbra{l}{j})
\end{equation}
which is equivalent to
\begin{equation}\label{eq005}
\hat{\Phi}_D \cdot \hat{\Phi} = \sum_{x,i,j}^N c_{i,j,x,x}\ketbra{x}{i}\otimes\ketbra{x}{j}.
\end{equation}
Jamio{\l}kowski representation \cite{BengtssonZyczkowski2008} of $\hat{\Phi}_D \cdot \hat{\Phi}$ is given by the form
\begin{equation}
J(\hat{\Phi}_D \cdot \hat{\Phi})=\sum_{x,i,j}^N c_{i,j,x,x}\ketbra{x}{x}\otimes\ketbra{i}{j}.
\end{equation}
From the property TP of quantum channels shows that $\tr_1 J(\hat{\Phi}_D \cdot \hat{\Phi}) = \1$ thus
\begin{equation}
\tr_1 J(\hat{\Phi}_D \cdot \hat{\Phi})=\sum_{i,j}^N \left(\sum_x^N c_{i,j,x,x}\right)\ketbra{i}{j} = \sum_{i,j} \delta_{i,j}\ketbra{i}{j}.
\end{equation}

\indent Submatrix determined by elements 
\begin{equation}\label{stochasticMatrix}
[c_{i,i,x,x}]_{i,x=0}^{N-1}
\end{equation}
is stochastic \cite{bruzda2009random}. Thus coin of the form $\Phi_D\cdot\Phi_C$ simulates the
classical queueing system.

\begin{exam}
Let's set the queue length dimension $D_Q=10$, it means that maximal length 
of queue can be $9$ and let's fix dimensions of registers used to
evaluating flow of jobs into the system in single time step $D_I=2$ and jobs
processed in single time step $D_O=2$. Initially, the probability distribution of the queue 
length is unimodal and has ``symmetric Bell'' shape. Next, the distribution begins to 
unify and after $49$ steps,  the probability distribution is close 
to the uniform distribution (Fig.~\ref{fig:ex2}).

\begin{figure}[h]
  \begin{center}
    \includegraphics[width=124mm]{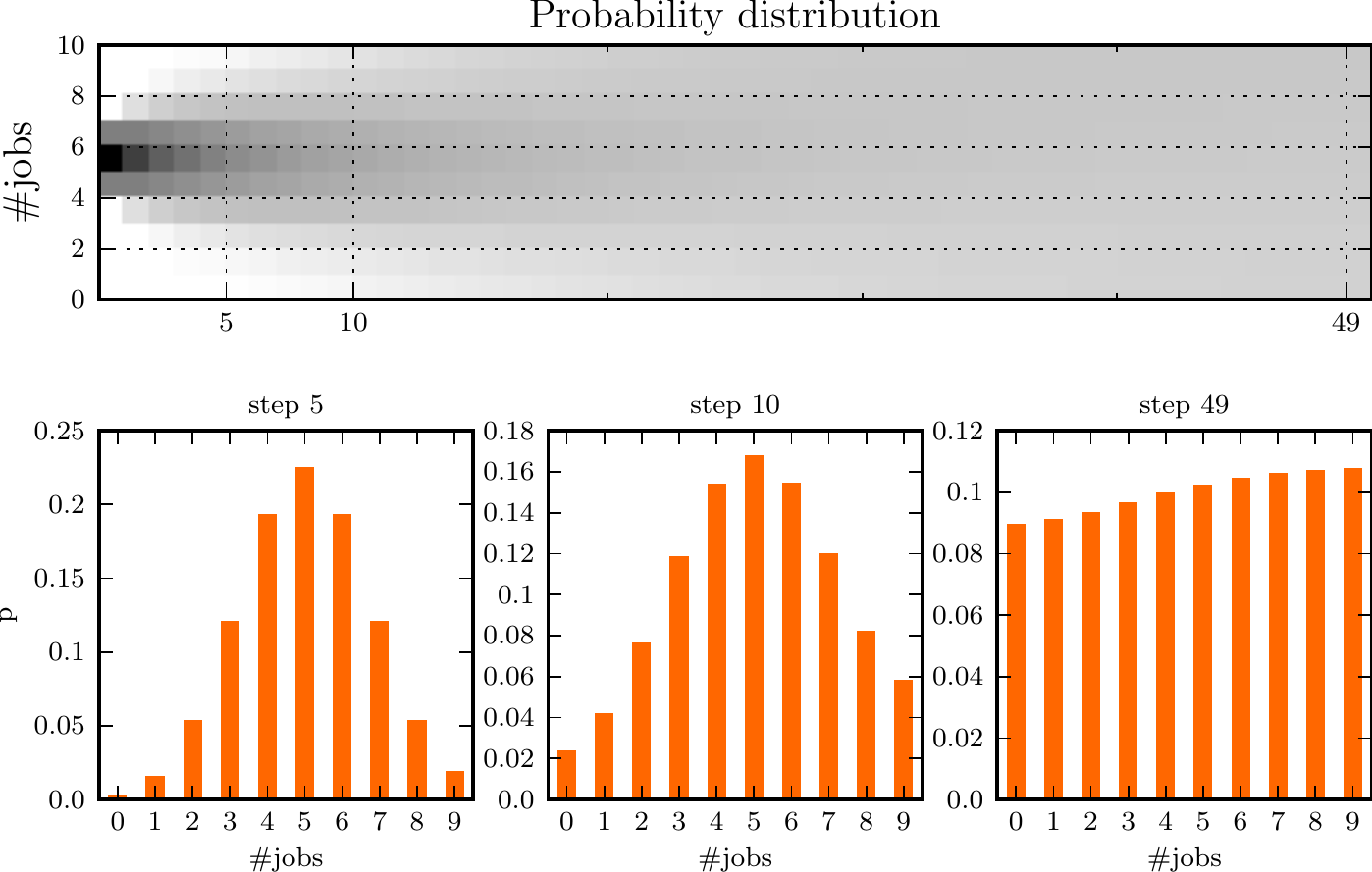}
  \end{center}
  \caption{The probability distribution of the queue length with classical coin. At the beginning, the probability distribution of the queue 
    length is unimodal. After $49$ steps the probability distribution is close to the uniform distribution.}
  \label{fig:ex2}
\end{figure}
\end{exam}

In classical Queueing Theory length of the queues can have different probability
distribution, e.g. Poisson distribution, Erlang distribution. Quantum queues can
also have various probability distribution of length depending on the stochastic matrix (\ref{stochasticMatrix}).







\section{The mean probability distribution of the queue length}
\label{sec:mean-distribution}

Consider queuing system performed by mapping $\Phi$ such that
$\lim_{t\to\infty}C\Phi^t$ is convergent, where $C$ is the partial trace and von
 Neumann measurement performed over subsystem $Q$. The probability distribution
of  the queue length is stationary and depend on the initial state $\rho$. Using
the Monte Carlo method it is possible to calculate the mean probability
distribution of the queue length for every initial state $\rho$. Consider the
set of random states $\Omega$ and a Hilbert-Schmidt measure $\mu$  discussed in
Subsection~\ref{randomState}, then the mean probability distribution of the
queue length over initial states from $\Omega$ is given by
\begin{equation}
\lim_{t\to\infty}\int_\Omega C\Phi^t(\rho) d\mu(\rho).
\end{equation}

\begin{exam}
Let the sizes of registers used to evaluating flow of jobs into system and jobs 
processed in single time step be $D_I=4$, $D_O=4$, respectively. Let's fix the register 
dimension of queue length $D_Q$ to $10$ in single time step and $\Omega$ as a set of
mixed random states. Quantum random walks are simulated using the Walsh-Hadamard,
Grover and $DFT$ coins. More precisely, in each case QRW are simulated by coin
operator given as
\begin{equation}
  \Phi_{U}(\rho)=(U^{I}\otimes U^{O}\otimes \1^Q) \rho (U^{I}\otimes U^{O}\otimes \1^Q)^\dagger,
\end{equation}
where $U$ is a one of the above coin. For the $600~000$ initial mixed random states obtained the following mean probability distributions of the queue length.

\paragraph{Case 1} In the first case, quantum random walks been implemented via a Walsh-Hadamrd coin $H_n = \frac{1}{\sqrt{2}}\left(\begin{smallmatrix}
1 & 1 \\
1 & - 1\\
\end{smallmatrix}\right)^{\otimes n}$. In this example $n$ is equal to $2$ and the mean probability distributions of the queue length is presented in Fig \ref{fig:WalshHadamard}.

\begin{figure}[h]
  \begin{center}
	\includegraphics[scale=1]{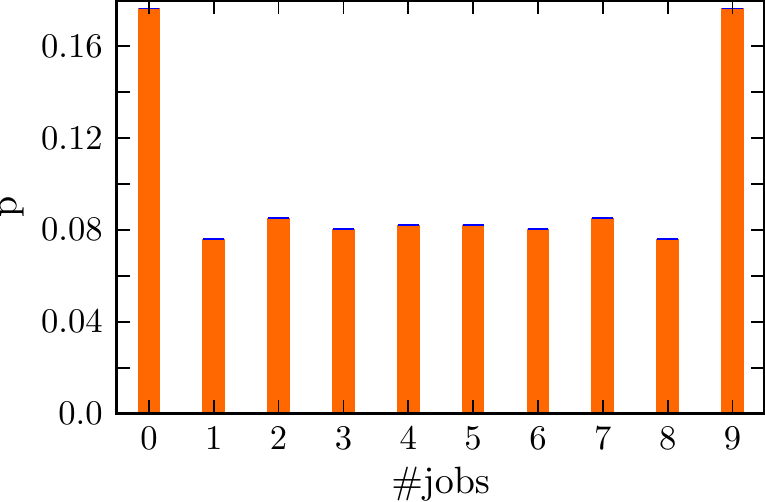}
  \end{center}
  \caption{The mean probability distribution of the queue length with Walsh-Hadamard coin.}
  \label{fig:WalshHadamard}
\end{figure}

\paragraph{Case 2} In the second case, QRW  are simulated using the Grover coin 
\begin{equation}
G=
\begin{pmatrix}
\frac{2}{d}-1 & \frac{2}{d} & \dots & \frac{2}{d} \\
\frac{2}{d} & \frac{2}{d}-1 & \dots & \frac{2}{d} \\
\vdots & \vdots & \ddots & \vdots \\
\frac{2}{d} & \frac{2}{d} & \dots & \frac{2}{d}-1 
\end{pmatrix},
\end{equation}
where in our case $d=4$ and results are shown in Fig. \ref{fig:Grover}.
\begin{figure}[h]
  \begin{center}
	\includegraphics[scale=1]{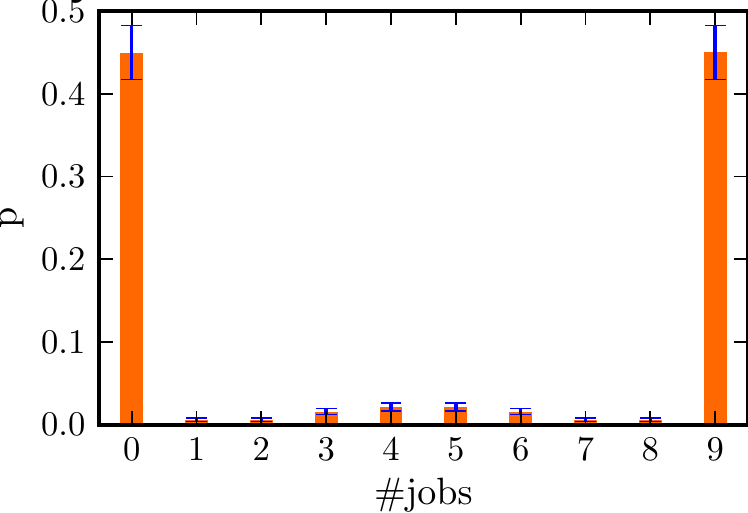}
  \end{center}
  \caption{The mean probability distribution of the queue length with Grover 
  coin. Error bars indicate plus minus one standard deviation.}
  \label{fig:Grover}
\end{figure}

\paragraph{Case 3} In the third case, quantum random walks are realized by DFT-coin
\begin{equation}
D=\frac{1}{\sqrt{d}}
\begin{pmatrix}
1 & 1 & 1 & \dots & 1 \\
1 & \omega & \omega^2 & \dots & \omega^{d-1} \\
\vdots & \vdots & \vdots & \ddots & \vdots\\
1 & \omega^{d-1} & \omega^{2(d-1)} & \dots & \omega^{(d-1)(d-1)}
\end{pmatrix},
\end{equation}
where $d=4$. For this coin, the mean probability distributions of the queue length is shown in Fig. \ref{fig:DFT}.

\begin{figure}[h]
  \begin{center}
	\includegraphics[scale=1]{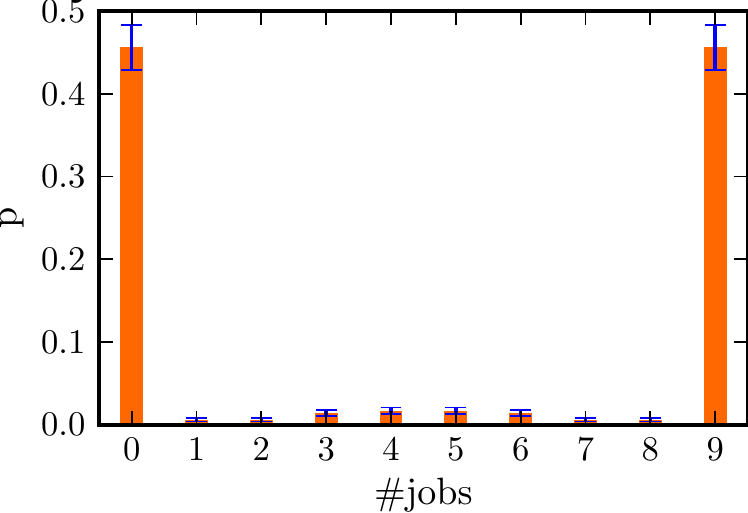}
  \end{center}
  \caption{The mean probability distribution of the queue length with DFT coin.
  Error bars indicate plus minus one standard deviation.}
  \label{fig:DFT}
\end{figure}

\end{exam}

The coins selection affects the probability distribution of the queue length. 
For this coins the standard deviations are relatively low so that the mean 
probability distribution of the queue length is a good estimator of queue length.  
While in case of Hadamard coin standard deviations are of the order of $10^{-12}$, 
thus initial states have little impact on queue length.

\section{Conclusions}
\label{sec:conclusions}
We have presented an extension of a simple queuing model based on Discrete Time 
Quantum Markov Chains using approch similar to Quantum Random Walks. We have 
shown that our model reduces to the classical one. We have proposed a 
methodology of analysis for quantum queuing models based on quantum channels 
and random mixed states.

In cases we studied numerically we observed that quantum queues models behave 
in a fundamentally different way than classical models of classical queues. 
In quantum models the length of the queue tends to quickly reach the barriers 
and stay in them, while classical balanced models tend to steady state in form 
of flat distribution.

The spectral structure of quantum models is much more complicated than that of 
classical ones. For most cases of queueing systems modelled using Discrete Time 
Markov Chain there exists steady state distribution. We have shown that for our 
models of quantum queues steady state distribution might exits but is dependent 
on the initial state of the modelled system. Therefore we proposed to analyse 
such systems using random initial mixed quantum state.

\section*{Acknowledgements} 
We acknowledge the financial support by the Polish Ministry of Science and
Higher Education under the grant number N N516 481840. We wish to thank Prof. T.
Czach\'orski, Dr. R.~Winiarczyk and Prof. J.~Klamka for helpful remarks and
inspiring discussions. Numerical calculations presented in this work were
performed on the \texttt{Leming} server of The Institute of Theoretical and
Applied Informatics, Polish Academy of Sciences.

\bibliographystyle{plain}
\bibliography{quantum_queues}
\end{document}